\documentclass[twocolumn,showpacs,preprintnumbers,amsmath,amssymb,superscriptaddress,floatfix,nofootinbib]{revtex4-1}
\usepackage{graphicx}
\usepackage{amsmath}
\usepackage{amsfonts}
\usepackage{amssymb}
\usepackage{color}
\usepackage[colorlinks, citecolor=blue,anchorcolor=red,menucolor=red, linkcolor=red,filecolor=red,runcolor=red,urlcolor=blue,frenchlinks=red]{hyperref}

\begin{document}

\title{Search for the $\Sigma^*$ state in $\Lambda^+_c \to \pi^+ \pi^0 \pi^-\Sigma^+$ decay by triangle singularity}

\author{Ju-Jun Xie} \email{xiejujun@impcas.ac.cn}
\affiliation{Institute of Modern Physics, Chinese Academy of
Sciences, Lanzhou 730000, China}

\author{Eulogio Oset}
\email{oset@ific.uv.es} \affiliation{Departamento de F\'isica
Te\'orica and IFIC, Centro Mixto Universidad de Valencia-CSIC,
Institutos de Investigac\'ion de Paterna, Aptdo. 22085, 46071
Valencia, Spain }

\date{\today}
\begin{abstract}

A $\Sigma^*$ resonance with spin-parity $J^P = 1/2^-$ and mass in
the vicinity of the $\bar{K}N$ threshold has been predicted in the
unitary chiral approach and inferred from the analysis of CLAS data
on the $\gamma p \to K^+ \pi^0 \Sigma^0$ reaction. In this work,
based on the dominant Cabibbo favored weak decay mechanism, we
perform a study of $\Lambda_c^+ \to \pi^+ \pi^0 \Sigma^*$ with the
possible $\Sigma^*$ state decaying into $\pi^- \Sigma^+$ through a
triangle diagram. This process is initiated by $ \Lambda_c^+ \to
\pi^+ \bar{K}^*N $, then the $\bar{K}^*$ decays into $\bar{K} \pi$
and $\bar{K} N$  produce the $\Sigma^*$ through a triangle loop
containing $\bar{K}^* N \bar{K}$ which develops a triangle
singularity. We show that the $\pi^- \Sigma^+$ state is generated
from final state interaction of $\bar{K}N$ in $S$-wave and isospin
$I=1$, and the $\Lambda_c^+ \to \pi^+ \pi^0 \pi^- \Sigma^+$ decay
can be used to study the possible $\Sigma^*$ state around the
$\bar{K}N$ threshold. The proposed decay mechanism can provide
valuable information on the nature of the $\Sigma^*$ resonance and
can in principle be tested by facilities such as LHCb, BelleII and
BESIII.
\end{abstract}

\maketitle

\section{Introduction}   \label{section-introduction}

Investigating low-lying excited states of $\Sigma(1193)$,
$\Sigma^*$, with isospin $I=1$ and strangeness $S=-1$ is one of the
important issues in hadronic
physics~\cite{Klempt:2009pi,Crede:2013sze,Tanabashi:2018oca}. The
$\Sigma^*$ states were mostly produced and studied in
antikaon-nucleon reactions, and the information on their properties
is still rather limited~\cite{Tanabashi:2018oca}. Based on the
chiral unitary approach, the low-lying $S=-1$ excited baryons were
studied by means of coupled channels in
Refs.~\cite{Kaiser:1995eg,Oset:1997it,Oller:2000fj,Oset:2001cn,Lutz:2001yb,GarciaRecio:2002td,Jido:2003cb,Oller:2006jw,Guo:2012vv,Hyodo:2011ur,Kamiya:2016jqc,Khemchandani:2018amu}.
In addition to the well reproduced properties of the
$\Lambda(1405)$, a possible resonance-like structure in $I=1$ around
the $\bar{K}N$ threshold was found in Ref.~\cite{Oller:2000fj} as a
bound state and in Ref.~\cite{Jido:2003cb} as a strong cusp effect.
Then it was further investigated in Ref.~\cite{Roca:2013cca} based
on the analysis of the experimental data on the $\gamma p \to K^+
\pi^{\pm}\Sigma^{\mp}$
reactions~\cite{Moriya:2013eb,Moriya:2013hwg}. Such a state near the
$\bar{K}N$ threshold is also discussed in
Refs.~\cite{Oller:2006jw,Guo:2012vv}, while in
Ref.~\cite{Khemchandani:2018amu}, a $\Sigma^*$ state is found with
mass around 1400 MeV, though it is not clear if it is related to one
or two poles in the complex plane. On the other hand, the effect of
this possible $\bar{K}N$ state with mass about $1430$ MeV in the
processes of $\chi_{c0}(1P) \to \pi \bar{\Sigma}\Sigma ~(\Lambda)$
decays was also studied in Refs.~\cite{Wang:2015qta,liu:2017efp}.

The nonleptonic weak decays of charmed baryons, particularly
$\Lambda^+_c$ decays into two mesons and one baryon, have shown a
great value to produce baryonic resonances with $S=-1$ and learn
about their
nature~\cite{Hyodo:2011js,Miyahara:2015cja,Xie:2016evi,Xie:2017xwx}.
In some cases the production rate is enhanced by the presence of a
triangle singularity (TS) in the reaction
mechanism~\cite{Xie:2017mbe,Dai:2018hqb}. The TS appears from a loop
diagram in the decay of a particle $1$ into two particles $2$ and
$3$ through the following process: at first the particle $1$ decays
into particles $A$ and $B$, and the particle $A$ subsequently decays
into particles $2$ and $C$, and finally the particles $B$ and $C$
merge and form the particle $3$ in the final state.

Following the work of Ref.~\cite{Dai:2018hqb}, in this paper, we
focus on the $\Lambda_c^+ \to \pi^+ \pi^0 \pi^-\Sigma^+$ decay
taking into account the $\pi^- \Sigma^+$ final state interaction
from the triangle diagram. We consider the external $W^+$ emission
diagram for the transition of $\Lambda_c^+$ into $\pi^+ \bar{K}^*
N$, which gives the main contribution to the $\Lambda_c^+ \to \pi^+
\pi^0 \pi^-\Sigma^+$ process~\cite{Dai:2018hqb}. Since we take into
account the external $W^+$ emission diagram, the $W^+$ produces the
$\pi^+$ in one vertex and in the other one includes a $c \to s$
transition. Then we have a $\pi^+$ and an $sud$ cluster, with the $ud$
diquark in $I=0$, because there these quarks are spectators. Hence
the $sud$ cluster hadronizes in $\bar{K}^* N$ in $I=0$, and, since
$\pi$ has isospin 1, the $\pi \Sigma$ system should be also in
isospin 1 to keep the isospin conserved. We will show that the
production of the $\bar{K} N$ state is enhanced by the TS in the
$\pi^0 \pi^- \Sigma^+$ mass distribution, and that a narrow peak or
cusp structure around the $\bar{K} N$ threshold in the $\pi\Sigma$
mass distribution appears. The observation of the TS in this process
would give further support to the existence of the $\bar{K} N$
resonance,~\footnote{In the following, we use $\Sigma^*(1430)$ to
denote the $\bar{K}N$ resonance.} and provide us better
understanding on the triangle singularity.

This article is organized as follows. In Sec.~\ref{Sec:Formalism},
we present the theoretical formalism for calculating the decay
amplitude of $\Lambda_c^+ \to \pi^+ \pi^0 \pi^-\Sigma^+$. Numerical
results and discussions are presented in Sec.~\ref{Sec:Results},
followed by a summary in the last section.


\section{Formalism}   \label{Sec:Formalism}

\begin{figure}[htbp]
\begin{center}
\includegraphics[scale=0.6]{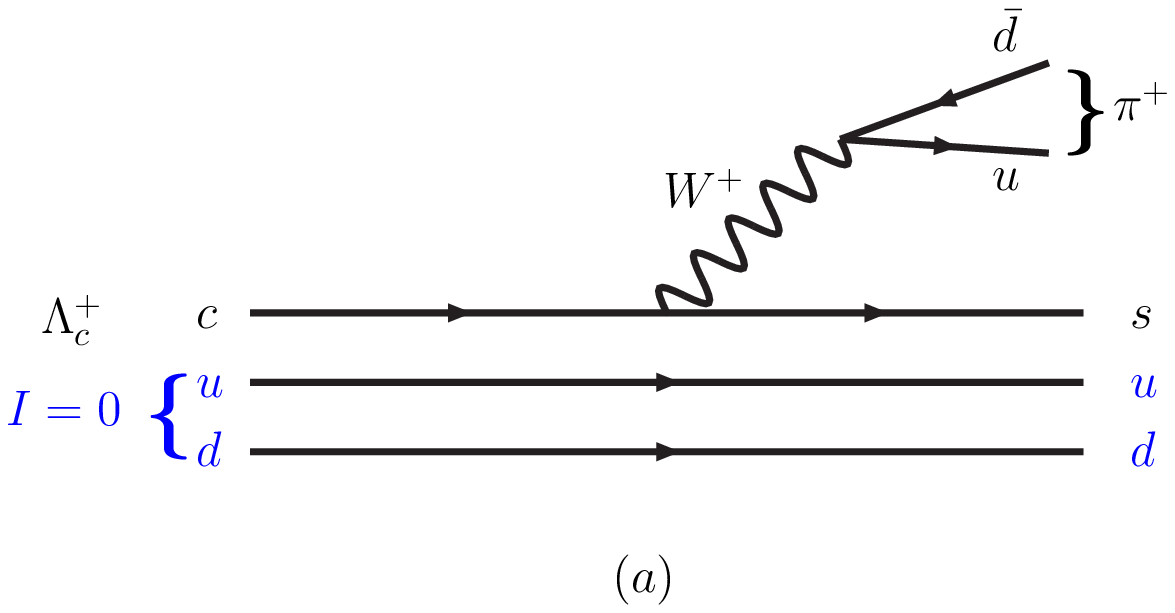}
\includegraphics[scale=0.6]{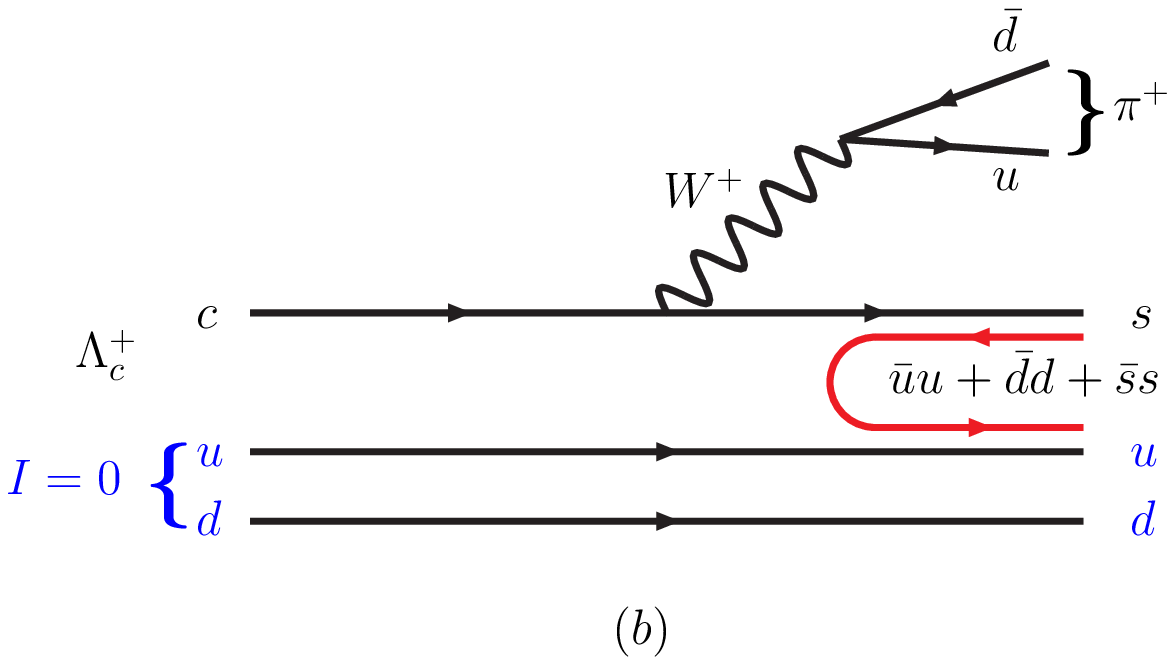}
\caption{(a) Quark level diagram for $\Lambda_c^+ \to \pi^+ sud$.
(b) Hadronization through $\bar{q}q$ creation with vacuum quantum
numbers.}\label{fig:feyndiag-quark}
\end{center}
\end{figure}

We consider the external emission mechanism of
Fig.~\ref{fig:feyndiag-quark}. The mechanism is Cabibbo favored. In
Fig.~\ref{fig:feyndiag-quark} (a) we see that the original $ud$
quarks of the $\Lambda^+_c$ are in $I=0$ and furthermore they are
spectators in the reaction, hence they continue to have $I=0$ in the
final $sud$ state. This state hadronizes to meson-baryon pairs after
creating a $\bar{q}q$ pair with the quantum numbers of the vacuum.
In our case we are interested in the $\bar{K}^*N$ production. The
insertion of $\bar{q}q$ can be done between any two quarks, but in
our case it must involve the strange quark. The reason is that we
want to have $\bar{K}^* N$ in $S$-wave, hence negative parity. Since
the $ud$ quarks are spectators and have positive parity, it must be
the strange quark that is produced in $L=1$. But we want it in its
ground state in the $\bar{K}^*$ after the hadronization, hence it
has to be involved in the hadronization process.

The explicit flavor combination of meson and baryon pairs of the
hadronization [see the process shown in
Fig.~\ref{fig:feyndiag-quark} (b)], $H$, is shown in
Ref.~\cite{Dai:2018hqb} and one finds,
\begin{eqnarray}
H = K^{*-}p + \bar{K}^{*0}n - \frac{\sqrt{6}}{3} \phi \Lambda,
\label{eq:H}
\end{eqnarray}
and one ignores the $\phi \Lambda$ component that has no role in the
TS mechanism. We can see that $H$ in Eq.~\eqref{eq:H} has $I=0$,
since $(\bar{K}^{*0},-K^{*-})$ is our isospin doublet. This
corresponds to the isospin of $s(ud)_{I=0}$ just after the weak
vertex, which is conserved after that.

\begin{figure*}[htbp]
\begin{center}
\includegraphics[scale=1.]{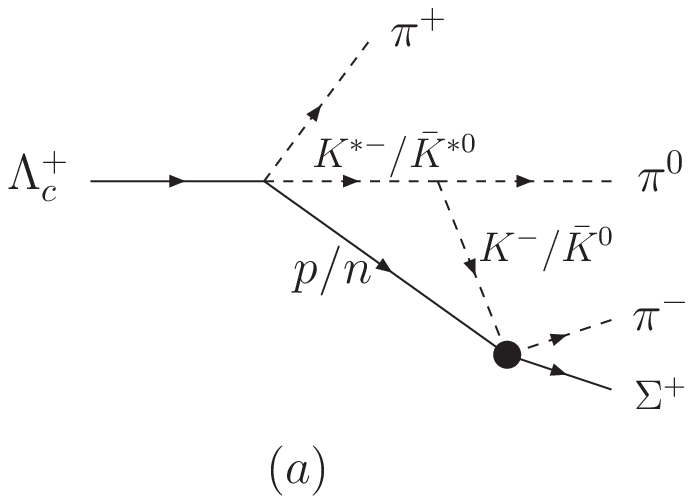} \hspace{2.cm}
\includegraphics[scale=1.]{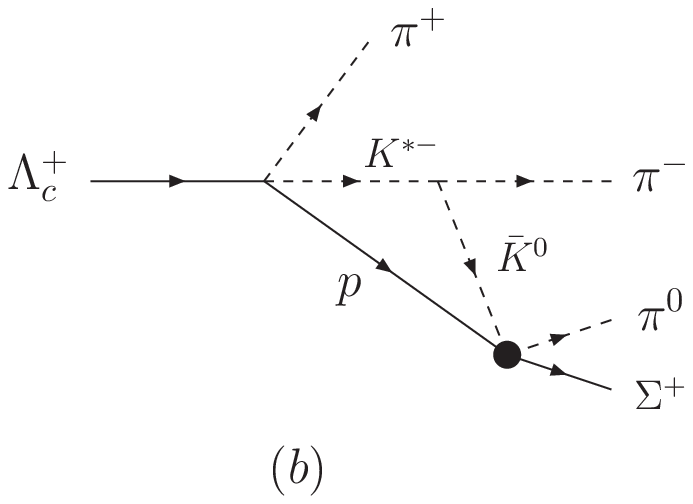}
\caption{Triangle diagrams for the decay of $\Lambda_c^+ \to \pi^+
\pi^0 \pi^-\Sigma^+$. (a) represents the final state interaction of
$\bar{K}N \to \pi^- \Sigma^+$, while (b) represents the final state
interaction of $\bar{K}^0 p \to \pi^0 \Sigma^+$. The black `dot'
stands for the vertex of the final state interaction of $\bar{K} N
\to \pi \Sigma$ in $S$-wave.}\label{fig:feyndiag}
\end{center}
\end{figure*}

Once $\pi^+ \bar{K}^*N$ is produced, the $\bar{K}^*$ decays to $\pi
\bar{K}$ and the $\bar{K}N$ interact to give $\pi \Sigma$. Since $\bar{K}^*N$ is in $I=0$, so must be the
$\pi \pi \Sigma$ system, which forces the $\pi \Sigma$ system, coming from the $\bar{K}N$ interaction, to have $I=1$ if there is
isospin conservation, as we shall assume here. This is shown in
Fig.~\ref{fig:feyndiag}.

By filtering the $I=1$ $\pi \Sigma$ system, the $\bar{K}N \to \pi
\Sigma$ amplitude incorporates the $\Sigma^*(1430)$ resonance and we
should see the signal of the state clearly in the $\pi \Sigma$ mass
distribution. In addition, as we shall see, the mechanism develops a
TS which enhances the production of the $\Sigma^*(1430)$ state. In
Fig.~\ref{fig:feyndiag} we have separated the two possible decay
modes: Fig.~\ref{fig:feyndiag} (a) shows the contribution from the
final state interaction of $\bar{K}^* N \to \pi^- \Sigma^+$, while
Fig.~\ref{fig:feyndiag} (b) stands for the transition of $\bar{K}^0
p \to \pi^0 \Sigma^+$.

We first consider the decay of $\Lambda_c^+ \to \pi^+ K^{*-} p$,
proceeding via $S$-wave, and we take the decay amplitude
as~\cite{Dai:2018hqb},
\begin{equation}\label{eq:A}
 t_{\Lambda_c^+ \to  \pi^+ K^{*-} p} = A {\vec\sigma} \cdot  {\vec\epsilon},
\end{equation}
with $A$ constant. Then we can easily obtain the branching ratio
$Br(\Lambda_c^+ \to  \pi^+ K^{*-} p)$, summing over the
$\bar{K}^{*-}$ polarizations, as
\begin{eqnarray}
&& Br(\Lambda_c^+ \to  \pi^+ K^{*-} p) = \nonumber
\\
&& \frac{3m_N|A|^2}{8\pi^3 M_{\Lambda_c^+} \Gamma_{\Lambda_c^+}}
\times \int^{M_{\Lambda_c^+} - m_{\pi^+}}_{m_{K^{*-}} + m_p }
p_{\pi^+} \tilde{p}_{K^{*-}} dM_{K^{*-} p},
\end{eqnarray}
where $p_{\pi^+}$ is the momentum of $\pi^+$ in the $\Lambda_c^+$
rest frame, and $\tilde{p}_{K^{*-}}$ is the momentum of
$\bar{K}^{*}$ in the $K^{*-} p$ rest frame with invariant mass
$M_{K^{*-} p}$,
\begin{eqnarray}
p_{\pi^+} &=& \frac{\lambda^{1/2} (M_{\Lambda_c^+}^2, m^2_{\pi}, M^2_{K^{*-} p})}{2 M_{\Lambda_c}},    \\
\tilde{p}_{K^{*-}} &=& \frac{\lambda^{1/2} (M^2_{K^{*-} p},
m^2_{K^{*-}}, m^2_{p})}{2M_{K^{*-} p}},
\end{eqnarray}
with $\lambda(x,y,z)$ the ordinary K{\"a}llen function.

By calculating the partial decay width of $\Lambda_c^+ \to \pi^+
K^{*-} p$, using the experimental branching ratio of $Br(\Lambda_c^+
\to \pi^+ K^{*-} p) = (1.4 \pm 0.5) \times 10^{-2}$ and
$\Gamma_{\Lambda^+_c} = (3.3 \pm 0.1) \times 10^{-9}$
MeV~\cite{Tanabashi:2018oca}, we can determine the value of the
constant $|A|^2$,
\begin{eqnarray}
|A|^2 = (3.9 \pm 1.4) \times 10^{-16} ~ {\rm MeV}^{-2},
\end{eqnarray}
where the error is taken from the experimental error in the
branching ratio of $Br(\Lambda_c^+ \to \pi^+ K^{*-} p)$. In view of
the weights in $H$ in Eq.~\eqref{eq:H}, the same value of $|A|^2$ is
used for the decay of $\Lambda_c^+ \to \pi^+ \bar{K}^{*0} n$.

Next, we write the total decay amplitude of $\Lambda_c^+ \to \pi^+
\pi^0 \pi^- \Sigma^+$ for those diagrams shown in
Fig.~\ref{fig:feyndiag},~\footnote{More details can be found in
Ref.~\cite{Dai:2018hqb}.}
\begin{equation}
t_{\rm total} = -\frac{Ag}{\sqrt{2}} \left ({\vec \sigma} \cdot
{\vec k}_a t^a_T {\cal M}^a  + \sqrt{2} {\vec \sigma} \cdot {\vec
k}_b t^b_T {\cal M}^b \right ), \label{eq:totalamplitude}
\end{equation}
where $\vec{k}_a$ and $\vec{k}_b$ are the momenta of the $\pi^0$ and
$\pi^-$ in the diagrams of Fig.~\ref{fig:feyndiag} (a) and (b),
respectively, calculated in the $\pi^0 \pi^- \Sigma^+$ rest frame,
and the coupling $g$ is given by $g = m_V/2f_\pi$ with $m_V = 780$
MeV and $f_\pi = 93$ MeV. In Eq.~\eqref{eq:totalamplitude}, ${\cal
M}^a$ and ${\cal M}^b$ are the two-body scattering amplitudes, which
depend on the invariant masses of $M_{\pi^- \Sigma^+}$ and $M_{\pi^0
\Sigma^+}$, respectively, and they have the explicitly forms as,
\begin{eqnarray}
{\cal M}^a &=&  t_{K^- p \to \pi^- \Sigma^+} - t_{\bar{K}^0 n \to \pi^- \Sigma^+}, \label{eq:ma} \\
{\cal M}^b &=&  t_{\bar{K}^0 p \to
 \pi^0 \Sigma^+}, \label{eq:mb}
\end{eqnarray}
where $t_{K^- p \to \pi^- \Sigma^+}$ and $t_{\bar{K}^0 n \to \pi^-
\Sigma^+}$ depend on $M_{\pi^- \Sigma^+}$, and $t_{\bar{K}^0 p \to
 \pi^0 \Sigma^+}$ depends on $M_{\pi^0 \Sigma^+}$. The factor
 $\sqrt{2}$ in Eq.~\eqref{eq:totalamplitude} and the minus sign in
 Eq.~\eqref{eq:ma} have their origin in the different $\bar{K}^* \to \pi
K$ vertices. Note that the two combinations, $K^- p - \bar{K}^0 n$ and $\bar{K}^0 p$ have $I=1$, as it should be.

In addition, we give explicitly the amplitude $t^a_T$ for the case
of $K^{*-}$, $p$ and $K^-$ in the triangle loop, as an example,
\begin{eqnarray}
&& t^a_T = \int \frac{d^3 q}{(2 \pi)^3} \frac{m_p}{2\omega_{K^{*-}}
\omega_p \omega_{K^-}}
\frac{1}{k^0_a-\omega_{K^-}-\omega_{K^{*-}}+i\frac{\Gamma_{K^{*-}}}{2}}
\nonumber \\
&& \times \frac{1}{P^0+\omega_p+\omega_{K^-}-k^0_a} \left(2+
\frac{\vec{q}
\cdot \vec{k}_a}{|\vec{k}_a|^2}\right)
\nonumber \\
&& \times \frac{P^0 \omega_p + k^0_a \omega_{K^-} -
(\omega_p+\omega_{K^-})(\omega_p+\omega_{K^-}+\omega_{K^{*-}})}{P^0-\omega_{K^{*-}}-\omega_p+i\frac{\Gamma_{K^{*-}}}{2}}
\nonumber \\
&&\times \frac{1}{P^0 - \omega_p -\omega_{K^-}-k^0_a+ i \epsilon} ,
\label{eq:ttriangle}
\end{eqnarray}
with  $P^0 = M_{\pi^- \pi^0 \Sigma^+}$ the invariant mass of the
final $\pi^0 \pi^- \Sigma^+$ system,
$\omega_p=\sqrt{|\vec{q}|^2+m_p^2}$,
$\omega_{K^-}=\sqrt{|\vec{q}+\vec{k}_a|^2+m_{K^-}^2}$, and
$\omega_{K^{*-}}=\sqrt{|\vec{q}|^2+m_{K^{*-}}^2}$. The energy
$k^0_a$ and momentum $|\vec{k}_a|$ of $\pi^0$ emitted from $K^{*-}$
are given by
\begin{eqnarray}
k^0_a &=& \frac{M^2_{\pi^- \pi^0 \Sigma^+} + m_{\pi^0}^2 -
M^2_{\pi^-
\Sigma^+}}{2 M_{\pi^- \pi^0 \Sigma^+}}, \\
|\vec{k}_a| &=& \sqrt{(k^{0})^2 - m^2_{\pi^0}}.
\end{eqnarray}
While, $k^{0}_b$, $|\vec{k}_b|$, and $t^b_T$ can easily be obtained
just applying the substitution to $k^0_a$, $|\vec{k}_a|$ and $t^a_T$
with $m_{\pi^0} \to m_{\pi^-}$ and $M_{\pi^- \Sigma^+} \to
M_{\pi^0\Sigma^+}$.

Then we obtain the final invariant masses distribution for four
particles in the final state,
\begin{eqnarray}
&& \frac{d^3 \Gamma}{d M_{\pi^0 \pi^- \Sigma^+} d M_{\pi^- \Sigma^+}
dM_{\pi^0 \Sigma^+}} = \frac{g^2|A|^2}{128 \pi^5}
\frac{m_{\Sigma^+}}{M_{\Lambda_c^+}} \widetilde{p}_{\pi^+} \nonumber \\
&& \times \frac{M_{\pi^- \Sigma^+}M_{\pi^0 \Sigma^+}}{ M_{\pi^0
\pi^- \Sigma^+}} \left( |\vec{k}_a|^2|t^a_T{\cal M}_a |^2 + 2|\vec{k}_b|^2|
 t^b_T{\cal M}_b|^2 \right. \nonumber \\
&& \left. + 2\sqrt{2} {\rm Re}[t^a_T{\cal M}_a (t^b_T{\cal M}_b)^*] \vec{k}_a \cdot \vec{k}_b \right ) , \label{eq:dGdM3}
\end{eqnarray}
with
\begin{eqnarray}
\widetilde{p}_{\pi^+} = \frac{\lambda^{1/2} (M_{\Lambda_c^+}^2,
M^2_{\pi^0 \pi^- \Sigma^+},m^2_{\pi^+})}{2 M_{\Lambda_c^+}
}.
\end{eqnarray}
In Eq.~\eqref{eq:dGdM3}, $\vec{k}_a \cdot \vec{k}_b$ is evaluated in terms of $M_{\pi^0 \pi^- \Sigma^+}$, $M_{\pi^- \Sigma^+}$, and $M_{\pi^0\Sigma^+}$,
\begin{eqnarray}
\vec{k}_a \cdot \vec{k}_b = \frac{m^2_{\pi^0} + m^2_{\pi^-} - M^2_{\pi^0 \pi^-} + 2k^0_a k^0_b}{2},
\end{eqnarray}
with
\begin{eqnarray}
 M^2_{\pi^0 \pi^-} &=& M^2_{\pi^0 \pi^- \Sigma^+} + m^2_{\pi^0} + m^2_{\pi^-} + m^2_{\Sigma^+} \nonumber \\
&&  - M^2_{\pi^- \Sigma^+} - M^2_{\pi^0 \Sigma^+}.
\end{eqnarray}

On the other hand, we have to regularize the integral in
Eq.~\eqref{eq:ttriangle}. In this work, we use the same cutoff of
the meson loop that is used to calculate $t_{K^- p \to
\pi^-\Sigma^+}$ with $\theta(q_{\text{max}}-|\vec{q}^*|)$, where
$\vec{q}^{\ *}$ is the $\vec{q}$ momentum in the $R$ rest frame (see
Ref.~\cite{Bayar:2016ftu} for more details).

\section{Numerical results} \label{Sec:Results}

\begin{figure}[htbp]
    \begin{center}
        \includegraphics[scale=0.45]{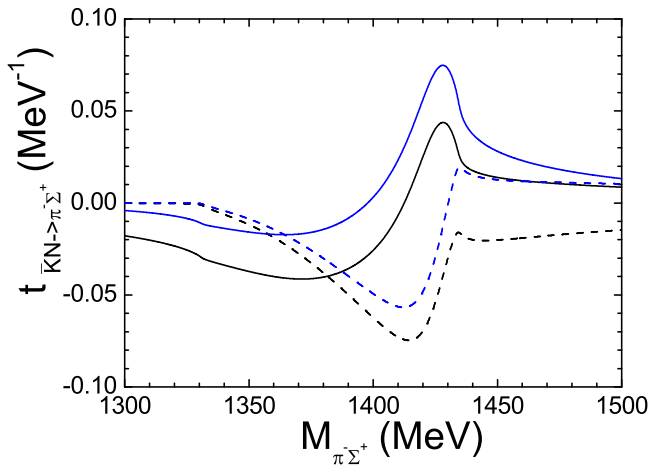}
        \caption{Transition amplitudes of $\bar{K}N \to \pi^- \Sigma^+$ in $S$-wave.}\label{Fig:tamplitude2body}
    \end{center}
\end{figure}

\begin{figure}[htbp]
    \begin{center}
        \includegraphics[scale=0.45]{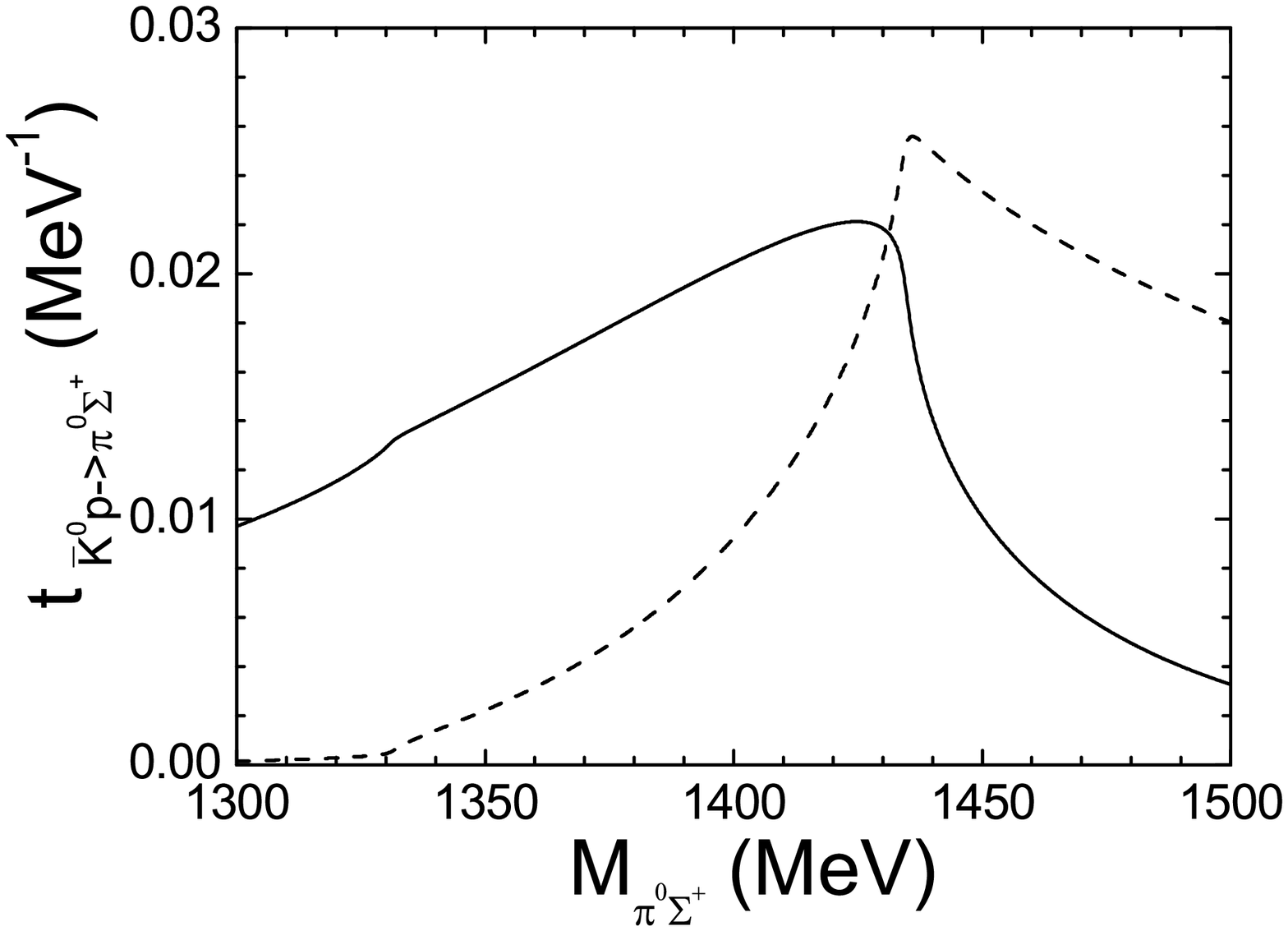}
        \caption{Transition amplitudes of $\bar{K}^0 p \to \pi^0 \Sigma^+$ in $S$-wave.}\label{Fig:tamplitude2bodynew}
    \end{center}
\end{figure}

In Figs.~\ref{Fig:tamplitude2body} and \ref{Fig:tamplitude2bodynew}
the two body transition amplitudes of $\bar{K}N \to \pi \Sigma$ are
shown. We plot the real and imaginary parts of those two body transition
amplitudes. In Fig.~\ref{Fig:tamplitude2body}, the blue curve and solid curve stand for the real parts of $t_{K^- p \to \pi^- \Sigma^+}$ and $t_{\bar{K}^0 n \to \pi^- \Sigma^+}$, respectively, while the blue-dashed curve and dashed curve stand for the imaginary parts of $t_{K^- p \to \pi^- \Sigma^+}$ and $t_{\bar{K}^0 n \to \pi^- \Sigma^+}$, respectively. In Fig.~\ref{Fig:tamplitude2bodynew}, the solid and dashed curves are the real and imaginary parts of $t_{\bar{K}^0 p \to \pi^0 \Sigma^+}$, respectively. It can be observed that the ${\rm Re}(t_{\bar{K}N \to
\pi^- \Sigma^+})$ have peaks around 1430 MeV, and $|{\rm
Im}(t_{\bar{K}N \to \pi^- \Sigma^+})|$ have peaks around 1420 MeV,
and there are bump structures for ${\rm Re}(t_{\bar{K}^0 p \to \pi^0
\Sigma^+})$ and ${\rm Im}(t_{\bar{K}^0 p \to \pi^0 \Sigma^+})$
around 1430 and 1440 MeV, respectively. These results, shown in
Figs.~\ref{Fig:tamplitude2body} and \ref{Fig:tamplitude2bodynew},
are obtained using the Bethe-Salpeter equation, with the tree level
potentials given in Ref.~\cite{Oset:1997it}. The loop functions for
the intermediate states are regularized using the cutoff method with
a cutoff of $630$ MeV. This parameter is also used to evaluate the
loop integral in the diagrams of Fig.~\ref{fig:feyndiag}.

\begin{figure}[htbp]
    \begin{center}
        \includegraphics[scale=0.5]{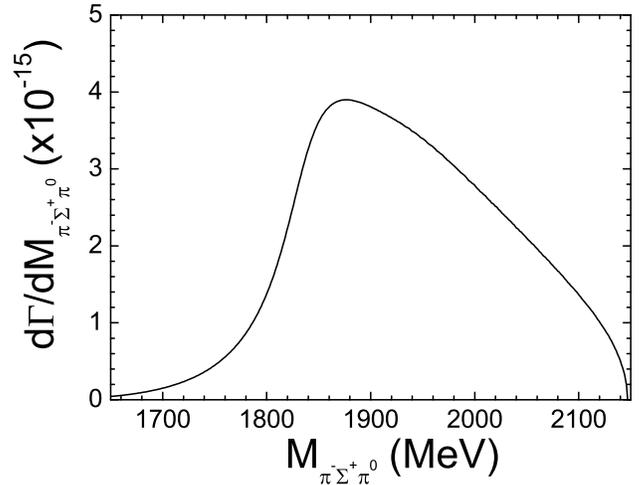}
        \vspace{-7.5cm}
        \caption{Invariant $\pi^0 \pi^- \Sigma^+$ mass distribution of $\Lambda^+_c \to \pi^+ \pi^0 \pi^- \Sigma^+$ decay.}\label{Fig:dgdM123}
    \end{center}
\end{figure}

With all the ingredients obtained above, one can easily obtain $d
\Gamma/d M_{\pi^0 \pi^- \Sigma^+}$ by integrating over $M_{\pi^-
\Sigma^+}$ and $M_{\pi^0 \Sigma^+}$ and using $|A|^2 = 3.9 \times
10^{-16}$ MeV$^{-2}$. We show the theoretical results of $d \Gamma/d
M_{\pi^0 \pi^- \Sigma^+}$ in Fig.~\ref{Fig:dgdM123}. We see a clear
bump structure of the invariant $\pi^0 \pi^- \Sigma^+$ mass
distribution around 1880 MeV for $\Lambda^+_c \to \pi^+ \pi^0 \pi^-
\Sigma^+$ decay, which is due to the triangle singularity of the
triangle diagrams as shown in Fig.~\ref{fig:feyndiag}.

\begin{figure}
    \begin{center}
        \includegraphics[scale=0.5]{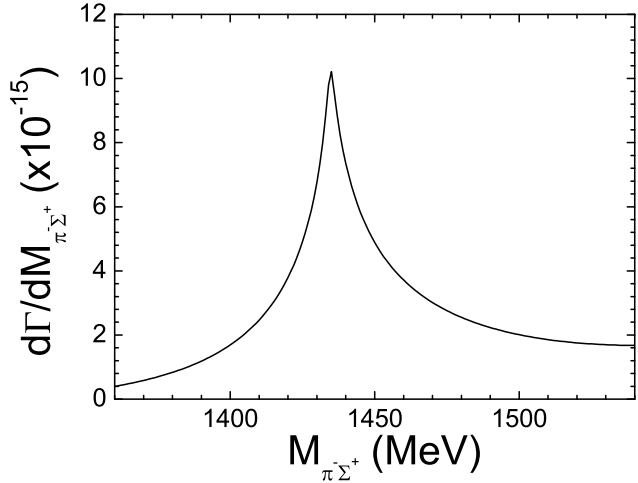}
        \vspace{-7.5cm}
        \caption{Invariant $\pi^- \Sigma^+$ mass distribution of $\Lambda^+_c \to \pi^+ \pi^0 \pi^- \Sigma^+$ decay.}\label{Fig:dgdM12}
    \end{center}
\end{figure}

On the other hand, by integrating over $M_{\pi^0 \pi^- \Sigma^+}$
and $M_{\pi^0 \Sigma^+}$, we obtain $d \Gamma/d M_{\pi^- \Sigma^+}$
which is shown in Fig.~\ref{Fig:dgdM12}. We see a really narrow peak
of the invariant $\pi^- \Sigma^+$ mass distribution around 1434 MeV
for $\Lambda^+_c \to \pi^+ \pi^0 \pi^- \Sigma^+$ decay, which is the
contribution from the $\bar{K}N$ resonance which is discussed above.
Similar results can be also obtained for $d \Gamma/d M_{\pi^0
\Sigma^+}$.

From these results shown in Figs.~\ref{Fig:dgdM123} and \ref{Fig:dgdM12}, one can easily obtain the branching ratio of $Br(\Lambda^+_c \to \pi^+ \pi^0 \pi^-\Sigma^+)$ is about $(3 \pm 1) \times 10^{-4}$, with the error that is taken from the error in $|A|^2$.

One should stress the most remarkable feature in the distributions
of Fig.~\ref{Fig:dgdM12}: the width of the $\Sigma^*(1430)$ produced
is about 10 MeV, remarkably smaller than the other $\Sigma^*$
resonances of 30 MeV or even bigger~\cite{Tanabashi:2018oca}. Yet,
as discussed in Ref.~\cite{Roca:2013cca} the peak corresponds to a
cusp at the $\bar{K}N$ mass threshold, however, very pronounced. The
dynamics of these cusps corresponds to a state nearly bound.
Theoretically, a small change in the parameters makes a pole appear.
This situation is very similar to the one of the $a_0(980)$
resonance, which both theoretically~\cite{Oller:1997ti} and
experimentally~\cite{Kornicer:2016axs} appears as a very pronounced
cusp.

\section{Conclusions} \label{sec:conc}

The triangle singularities have recently shown to be very important
in many hadronic decays. In this work we provide the first
evaluation of the $\Sigma^*(1430)$ production in the decay of
$\Lambda^+_c \to \pi^+ \pi^0 \pi^- \Sigma^+$. The decay mechanism
for the production is given by a first decay of the $\Lambda_c^+$
into $\pi^+ \bar K^* N $, then the $\bar{K}^*$ decays into $\bar K
\pi $ and the $\bar K N$ merge to produce the $\Sigma^*(1430)$
through both the final state interaction of $\bar{K} N \to \pi
\Sigma$ transition and a triangle loop containing $\bar{K}^* N \bar
K$, which develops a singularity of the invariant mass of $\pi^0
\pi^- \Sigma^+$ system around 1880 MeV.

It is found that a narrow peak, of the order of 10 MeV, tied to the
$\Sigma^*(1430)$ state appears in the final $\pi^- \Sigma^+$ mass
spectrum at the energy around the $\bar K N$ mass threshold of 1434
MeV. The line shape obtained here is intimately tied to the nature
of the $\Sigma^*(1430)$ as a dynamically generated resonance from
the meson baryon interaction, and shows up as a cusp structure. The
theoretical calculations done here together with the experimental measurements would thus bring valuable information
on the nature of this resonance. Corresponding
experimental measurements could in principle be done by
BESIII~\cite{Ablikim:2015flg} and BelleII~\cite{Yang:2015ytm}
Collaborations. In this sense the branching ratios obtained here for this $\Sigma^*(1430)$ signal are of the order of $10^{-4}$, which are well within the measurable range in these facilities. In view of that, the measurements of the $\Lambda^+_c \to \pi^+\pi^0
\pi^-\Sigma^+$ decay is strongly encouraged. Besides, the mechanism studied in this work contributes also to the processes of $\Lambda^+_c \to \pi^+ \pi^0 \pi^0 \Lambda$ and $\Lambda^+_c \to \pi^+ \pi^+ \pi^- \Lambda$, and thus these processes are also very interesting and can be measured by future experiments.

\vspace{1.cm}

\section*{Acknowledgements}

This work is partly supported by the National Natural Science
Foundation of China under Grant Nos. 11475227 and 11735003, and by
the Youth Innovation Promotion Association CAS (No. 2016367). It is
also partly supported by the Spanish Ministerio de Economia y
Competitividad and European FEDER funds under the contract number
FIS2011- 28853-C02-01, FIS2011- 28853-C02-02, FIS2014-57026- REDT,
FIS2014-51948-C2- 1-P, and FIS2014-51948-C2- 2-P, and the
Generalitat Valenciana in the program Prometeo II-2014/068 (EO).

\end{document}